\documentclass[reprint,twocolumn,secnumarabic,amssymb, nobibnotes, aps,prb]{revtex4-2}

\usepackage{amsmath,amssymb,pifont}
\usepackage{color}
\usepackage{graphicx}
\usepackage{pstool}
\usepackage{makecell}
\usepackage{amsfonts} 
\usepackage{dblfloatfix} 

\newcommand{\ve}[1]{\mathbf{#1}}
\newcommand{\te}[1]{\overline{\overline{#1}}}

\newcommand{\B}[1]{#1}

\usepackage{threeparttable}
\usepackage{amsfonts}
\usepackage{multirow}
\usepackage{float}

\usepackage{graphicx,bm}
\usepackage[usenames,dvipsnames]{xcolor}
\usepackage{subfig}
\usepackage{cancel}
\usepackage{pstool,pstricks}

\newcolumntype{N}{@{}m{0pt}@{}}

\newcounter{tempEquationCounter}
\newcounter{thisEquationNumber}

\begin{document}

\title{Extension of Lorentz Reciprocity and Poynting Theorems \\for Spatially Dispersive Media with Quadrupolar Responses}

\author{Karim~Achouri and Olivier J. F. Martin}

\email[E-mail: ]{karim.achouri@epfl.ch, olivier.martin@epfl.ch}
\affiliation{ Nanophotonics and Metrology Laboratory, Department
	of Microengineering, $\acute{E}$cole Polytechnique F$\acute{{e}}$d$\acute{{e}}$rale de Lausanne, Route Cantonale, 1015 Lausanne, Switzerland.}

\begin{abstract}
We provide a self-consistent extension of the Lorentz reciprocity theorem and the Poynting theorem for media possessing electric and magnetic dipolar and quadrupolar responses related to electric and magnetic fields and field gradients \B{thus corresponding to weak spatial dispersion}. Using these two theorems, we respectively deduce the conditions of reciprocity and gainlessness and losslessness that apply to the various tensors mediating the interactions of these multipole moments and the associated fields and field gradients. We expect that these conditions will play an essential role in developing advanced metamaterial modeling techniques that include quadrupolar and spatially dispersive responses. 
\end{abstract}

\maketitle

\section{Introduction}

The expansion of the electromagnetic fields induced by charges and current distributions in terms of multipolar moments plays an essential role in studying the interactions of light with matter~\cite{raabMultipoleTheoryElectromagnetism2005,agranovich2013crystal}. It has been widely used, for instance, in the electromagnetic characterizations of cubic and nonmagnetic crystals~\cite{LightPropagationCubic1990,grahamMultipoleMomentsMaxwell1992,grahamMagneticEffectsAntiferromagnetic1992} and the studies of optical effects such as optical activity and circular birefringence in chiral media~\cite{raab1994eigenvalue}. Historically, these studies have shown that, while dipolar approximations may be sufficient in some cases, it is often necessary to extend the multipolar expansion to higher order terms, such as quadrupoles or even octupoles, in order to properly assess the electromagnetic properties of the considered structures and model their optical response~\cite{LightPropagationCubic1990,grahamMultipoleMomentsMaxwell1992,grahamMagneticEffectsAntiferromagnetic1992,raab1994eigenvalue}.

In more recent years, we have witnessed the advances in the field of metamaterials and metasurfaces research that have led to a myriad of concepts and applications~\cite{landy2008perfect,schurig2006metamaterial,kildishev2013planar,yu2014flat,chenHuygensMetasurfacesMicrowaves2018,achouri2020electromagnetic}. In the case of metasurfaces, most of the design and modeling techniques that pertain to their synthesis and analysis are almost entirely based on dipolar approximations, since the scattering particles that compose them are small enough compared to the wavelength such that dipolar approximations are sufficient to model their responses~\cite{6477089,achouri2014general,epstein2016arbitrary}. However, the extreme number of degrees of freedom enabled by the ability to engineer the metasurface scattering particles may be leveraged to induce non-negligible higher order multipolar components for additional field control capabilities, the necessity to control the angular scattering response of metasurfaces for analog signal processing and the fact that dielectric metasurfaces are composed of dielectric resonators that are electrically large enough to exhibit at least electric quadrupolar responses prompt the need to improve the existing modeling techniques by including higher order multipolar components~\cite{petschulatUnderstandingElectricMagnetic2010,muhligMultipoleAnalysisMetaatoms2011,varault2013multipolar,silveirinhaBoundaryConditionsQuadrupolar2014,bernalarangoUnderpinningHybridizationIntuition2014,choContributionElectricQuadrupole2008,poutrina2014multipole,babichevaMetasurfacesElectricQuadrupole2018,momeni2019generalized,angularAchouri2020}.

However, the theoretical expressions derived in~\cite{raabMultipoleTheoryElectromagnetism2005,agranovich2013crystal,LightPropagationCubic1990,grahamMultipoleMomentsMaxwell1992,grahamMagneticEffectsAntiferromagnetic1992,simovski2018composite} have been limited to rather few multipole moments and related field interactions. \B{Moreover, in these studies, the permutation symmetries that are associated to quadrupolar components, and which correspond to reciprocity conditions, have only been derived based on quantum mechanical concepts}~\cite{raabMultipoleTheoryElectromagnetism2005,agranovich2013crystal,LightPropagationCubic1990,grahamMultipoleMomentsMaxwell1992,grahamMagneticEffectsAntiferromagnetic1992,simovski2018composite}.

In this work, we shall overcome these limitations by providing a multipolar expansion that includes electric and magnetic dipolar and quadrupolar moments with components related to the electric and magnetic fields as well as their first order derivatives. Additionally, we also provide the permutation symmetries that are associated to reciprocity and energy conservation based on a purely electromagnetic derivation of the Lorentz reciprocity theorem and the Poynting theorem. 

For self-consistency, we shall first review the fundamental concepts of the theory of multipoles and that of spatial dispersion in Secs.~\ref{sec:multi} and~\ref{sec:SD}, respectively. Next, we derive the Lorentz reciprocity theorem in Sec.~\ref{sec:lorentz} from which we derive the associated conditions of reciprocity. Then, in Sec.~\ref{sec:poyn}, we derive the Poynting theorem and deduce the conditions of gainlessness and losslessness, and finally conclude in Sec.~\ref{sec:concl}.

\section{Theoretical Background}

\subsection{Theory of Multipoles}
\label{sec:multi}

The electric and magnetic responses of a material or a medium are conventionally expressed in terms of constitutive relations~\cite{jackson_classical_1999,kong1986electromagnetic}. \B{In this work, we will consider that these constitutive relations are given by $\ve{D}$ and $\ve{B}$ as functions of $\ve{E}$ and $\ve{H}$, as it is particularly well-suited for the study of spatially dispersive metamaterials~\cite{capolino2009theory,simovski2018composite,varault2013multipolar}}. These relations are generally derived from a multipolar expansion of the current density induced in the material~\cite{jackson_classical_1999,papas2014theory}. For completeness, we next provide a brief derivation of this multipolar expansion. \B{We emphasize that several fundamental concepts related to the theory of multipoles are purposefully omitted in the upcoming discussion for briefness and we direct the interested readers to the extensive explanations provided in~\cite{raabMultipoleTheoryElectromagnetism2005,capolino2009theory,papas2014theory,simovski2018composite}.}

The typical procedure to expand the electric current density $\ve{J}$ in terms of multipoles is to start from the definition of the vector potential $\ve{A}$ from which the magnetic field is defined as $\ve{B} =\nabla\times\ve{A}$ and that is given by~\cite{jackson_classical_1999,papas2014theory}
\begin{equation}
\label{eq:Aint}
\ve{A}(\ve{r}) = \frac{\mu_0}{4\pi} \int \ve{J}(\ve{r'}) \frac{e^{-jk|\ve{r} - \ve{r'}|}}{|\ve{r} - \ve{r'}|}dV',
\end{equation}
where $k=2\pi/\lambda_0$ and the time-harmonic $e^{j\omega t}$ is assumed but omitted throughout~\footnote{\B{We emphasize that using $e^{j\omega t}$ or $e^{-j\omega t}$ does not change any of the fundamental reciprocity and gainless/lossless relations that are ultimately obtained throughout.}}. Using a Taylor expansion~\footnote{We use the following definition of the Taylor expansion: $f(\ve{r} + \ve{a}) = \sum_{n=0}^\infty \frac{1}{n!} (\ve{a}\cdot\nabla)^n f(\ve{r})$~\cite{papas2014theory}.}, the free-space Green function in~\eqref{eq:Aint} is expanded into
\begin{equation}
	\label{eq:GreenExp}
	\frac{e^{-jk|\ve{r} - \ve{r'}|}}{|\ve{r} - \ve{r'}|} = \sum_{n=0}^\infty \frac{1}{n!} (-\ve{r'}\cdot\nabla)^n \frac{e^{-jkr}}{r}.
\end{equation}
Substituting~\eqref{eq:GreenExp}, truncated at $n=2$, into~\eqref{eq:Aint} yields~\cite{papas2014theory}
\begin{equation}
	\label{eq:Aexp}
	\begin{split}
		\ve{A}(\ve{r}) = \frac{\mu_0}{4\pi}  \bigg[&\left(\int \ve{J}(\ve{r'})~dV' \right) -\left(\int \ve{J}(\ve{r'})\ve{r'}~dV' \right)\cdot\nabla+\\
		 &\qquad\frac{1}{2}\left(\int \ve{J}(\ve{r'})\ve{r'}\ve{r'}~dV' \right):\nabla\nabla\bigg]\frac{e^{-jkr}}{r}.
	\end{split}
\end{equation}
We next highlight the fact that the dyadic defined by $\ve{J}(\ve{r'})\ve{r'}$ in~\eqref{eq:Aexp} may be decomposed into symmetric and antisymmetric parts respectively as
\begin{equation}
	\label{eq:splitJ}
	\ve{J}(\ve{r'})\ve{r'} = \frac{1}{2}\left[\ve{J}(\ve{r'})\ve{r'} + \ve{r'}\ve{J}(\ve{r'})\right] + \frac{1}{2}\left[\ve{J}(\ve{r'})\ve{r'} - \ve{r'}\ve{J}(\ve{r'})\right],
\end{equation}
which plays an important role in simplifying~\eqref{eq:Aexp} for the following steps. 

\B{The integrals in brackets in~\eqref{eq:Aexp} can now be individually associated to specific multipole moments. By using~\eqref{eq:splitJ}, we thus define the electric dipole $\ve{p}$, the magnetic dipole $\ve{m}$, the electric quadrupole $\te{q}$ and the magnetic quadrupole $\te{s}$ as~\cite{grahamMultipoleMomentsMaxwell1992,simovski2018composite}}
\B{
\begin{subequations}
	\label{eq:mom}
\begin{align}
	p_i &= \frac{1}{j\omega}\int J_i~ dV,\\
	m_{i} &= \frac{1}{2}\int \left(\ve{r}\times\ve{J}\right)_i dV,	\\
	q_{ij}&= \frac{1}{j\omega}\int J_ir_j + J_jr_i~dV,\label{eq:Qij}	\\
	s_{ij} &= \frac{2}{3}\int \left(\ve{r}\times\ve{J}\right)_i r_j ~dV,
\end{align}	
\end{subequations}
}
where the Einstein summation convention over repeated indices is assumed. \B{Note that the term $\ve{J}(\ve{r'})\ve{r'}\ve{r'}$ in~\eqref{eq:Aexp} is conventionally split into the magnetic quadrupole moment $s_{ij}$ and into the electric octupole moment, which is intentionally omitted here. }

\B{We are next interested in the case of a bulk homogeneous medium assumed to be of infinite extent that, when excited by an electromagnetic field, exhibits a given distribution of current density. This bulk medium may be decomposed into deeply subwavelength subvolumes~\footnote{In the case where the bulk medium is a metamaterial, these subvolumes would correspond to the metamaterial unit cells.}, of volume $V_\text{sub}$, for which we may individually compute the corresponding multipole moments by evaluating the integrals in~\eqref{eq:mom} over $V_\text{sub}$.
}	

\B{We may now express the distribution of current density of the bulk medium in terms of the multipole moments as~\cite{raabMultipoleTheoryElectromagnetism2005,capolino2009theory,papas2014theory,simovski2018composite}
\begin{equation}
	\label{eq:Jmulti}
	\ve{J} = j\omega\ve{P} + \nabla\times\ve{M} - \frac{j\omega}{2}\nabla\cdot\te{Q}  - \frac{1}{2}\nabla\times(\nabla\cdot\te{S}),
\end{equation}
where $\ve{P}=\ve{p}/V_\text{sub}$, $\ve{M}=\ve{m}/V_\text{sub}$, $\te{Q}=\te{q}/V_\text{sub}$ and \mbox{$\te{S}=\te{s}/V_\text{sub}$} are spatially varying functions corresponding to dipolar and quadrupolar densities.}

The constitutive relations of that medium may now be derived by considering the frequency-domain Maxwell equations, expressed in terms of the fundamental fields $\ve{E}$ and $\ve{B}$~\cite{grahamMultipoleMomentsMaxwell1992,jackson_classical_1999}, given by
\begin{subequations}
	\label{eq:MEQF}
	\begin{align}
		\nabla\times\ve{E} &= -j\omega\ve{B},\label{eq:MEQF1}\\
		\nabla\times\ve{B} &= \mu_0\left(\ve{J}+j\omega\epsilon_0\ve{E}\right).\label{eq:MEQF2}
	\end{align}
\end{subequations}
Substituting~\eqref{eq:Jmulti} into~\eqref{eq:MEQF2} and re-arranging the terms, yields
\B{
\begin{equation}
	\label{eq:MEQ02}
	\nabla\times\left(\mu_0^{-1}\ve{B}-\ve{M} + \frac{1}{2}\nabla\cdot\te{S}\right) = j\omega\left(\epsilon_0\ve{E} + \ve{P} - \frac{1}{2}\nabla\cdot\te{Q}\right).
\end{equation}
}
We may now associate the first bracket in~\eqref{eq:MEQ02} to $\ve{H}$ and the second one to $\ve{D}$ as
\B{
\begin{subequations}
	\label{eq:DBPMQS00}
	\begin{align}
		\label{eq:DBPMQS001}\ve{D} &= \epsilon_0\ve{E} + \ve{P} - \frac{1}{2}\nabla\cdot\te{Q},\\
		\label{eq:DBPMQS002}\ve{H} &= \mu_0^{-1}\ve{B}-\ve{M} + \frac{1}{2}\nabla\cdot\te{S}.
	\end{align}
\end{subequations}
}
Since, in this work, we are interested in expressing the constitutive relations in terms of $\ve{D}$ and $\ve{B}$, \B{we deduce from~\eqref{eq:DBPMQS00} that the constitutive relations may be alternatively expressed as} 
\B{
\begin{subequations}
	\label{eq:DBPMQS0}
	\begin{align}
		\label{eq:DBPMQS01}\ve{D} &= \epsilon_0\ve{E} + \ve{P} - \frac{1}{2}\nabla\cdot\te{Q},\\
		\label{eq:DBPMQS02}\ve{B} &= \mu_0\left(\ve{H} + \ve{M}- \frac{1}{2}\nabla\cdot\te{S}\right),
	\end{align}
\end{subequations}
}
which we will now use as the de facto constitutive relations for the upcoming derivations.

\subsection{Spatial Dispersion}
\label{sec:SD}

Spatial dispersion describes the spatially nonlocal nature of electromagnetic material responses~\cite{serdkov2001electromagnetics,simovski2018composite}. It implies, for instance, that the electric current density $\ve{J}$ may be expressed as the convolution of an exciting electric field $\ve{E}$ with the current response of the material $\te{K}$ as~\cite{serdkov2001electromagnetics,simovski2018composite}
\begin{equation}
	\label{eq:WSDJ}
	\ve{J}(\ve{r}) = \int \te{K}(\ve{r}-\ve{r'})\cdot\ve{E}(\ve{r'})~dV'.
\end{equation}
We shall now approximate this expression by expanding the electric field as~\footnote{We use the three first terms of the Taylor expansion defined as $f(\ve{r'}) = \sum_{n=0}^\infty \frac{1}{n!} \left[(\ve{r'} -\ve{r} )\cdot\nabla\right]^n f(\ve{r})$.}
\begin{equation}
	\label{eq:Taylor2E}
	\ve{E}(\ve{r'}) = \ve{E}(\ve{r}) +  (\ve{r'} -\ve{r} )\cdot\nabla \ve{E}(\ve{r}) + \frac{1}{2} \left[(\ve{r'} -\ve{r} )\cdot\nabla\right]^2 \ve{E}(\ve{r}),
\end{equation}
which, upon insertion in~\eqref{eq:WSDJ}, yields~\cite{raabMultipoleTheoryElectromagnetism2005,capolino2009theory,papas2014theory,simovski2018composite}
\begin{equation}
	\label{eq:Jexp}
	J_i = b_{ij}E_j + b_{ijk}\nabla_kE_j + b_{ijkl}\nabla_l\nabla_kE_j,
\end{equation}
\B{where the rank-2 tensor $b_{ij} = \int \te{K}(\ve{r}-\ve{r'})dV'$, the rank-3 tensor $b_{ijk}=\int \te{K}(\ve{r}-\ve{r'})(\ve{r'} -\ve{r} )dV'$, and so on. Expression~\eqref{eq:Jexp} shows that the induced current is non-locally related to the exciting electric field via its spatial derivatives, which is the origin of spatial dispersion. This ultimately leads to material parameters that depend on the direction of wave propagation~\cite{capolino2009theory,simovski2018composite}.}

It is particularly important to note that the terms related to the derivatives of the electric field in~\eqref{eq:Jexp} may be transformed to make the magnetic field $\ve{H}$ appear. Indeed, consider, for instance, the second term on the right-hand side of~\eqref{eq:Jexp}, which may be split into symmetric and antisymmetric parts as $b_{ijk} = b_{ijk}^\text{sym}+b_{ijk}^\text{asym}$. Then, considering that an antisymmetric third-rank tensor may be represented by a second-rank pseudotensor implying the dual quantity $b_{ijk}^\text{asym}\propto\varepsilon_{ljk}g_{il}$~\cite{arfken1999mathematical}, where $\varepsilon_{ijk}$ is the Levi-Civita symbol and $g_{ij}$ is a second-rank tensor, we obtain
\begin{equation}
	\label{eq:bsplit}
	\begin{split}
			b_{ijk}\nabla_kE_j &= \left(b_{ijk}^\text{sym}+b_{ijk}^\text{asym}\right)\nabla_kE_j,\\
			&=\left(b_{ijk}^\text{sym} + \frac{j}{\omega\mu_0}\varepsilon_{ljk}g_{il}\right)\nabla_kE_j,\\
			&=b_{ijk}^\text{sym}\nabla_kE_j + g_{ij}H_j,
	\end{split}
\end{equation}
where we have used the fact that in vacuum~\eqref{eq:MEQF1} may, using $\ve{B}=\mu_0 \ve{H}$, be written as $\varepsilon_{ijk}\nabla_kE_j=-j\omega\mu_0 H_i$. Relation~\eqref{eq:bsplit} shows that the dependence of $\ve{J}$ on the magnetic field is on the same order as its dependence on the first order derivative of the electric field. Similarly, the last term in~\eqref{eq:Jexp} may be transformed into two terms, one related to the first derivative of $\ve{H}$ and one to the second derivative of $\ve{E}$~\cite{simovski2018composite}.

From a general perspective, the convolution~\eqref{eq:WSDJ} and its expansion~\eqref{eq:Jexp} may also be performed for any of the multipole moments in~\eqref{eq:mom}~\cite{varault2013multipolar,simovski2018composite}. Accordingly, we now expand these quantities in terms of the electric field, the magnetic field and their first order derivatives, which leads to
%
%
%
\begin{equation}
	\label{eq:PMQS}
	\begin{pmatrix}
		P_i\\
		M_i\\
		Q_{il}\\
		S_{il}
	\end{pmatrix}=
	\te{\chi}\cdot
	\begin{pmatrix}
	E_j\\
	H_j\\
	\nabla_k E_j\\
	\nabla_k H_j
\end{pmatrix},
\end{equation}
where the hypersusceptibility tensor $\te{\chi}$ is given by
\begin{equation}
	\label{eq:chi}
		\begin{pmatrix}
		\epsilon_0\chi_{\text{ee},ij} & \frac{1}{c_0}\chi_{\text{em},ij} & \frac{\epsilon_0}{2k_0}\chi'_{\text{ee},ijk} & \frac{1}{2c_0k_0}\chi'_{\text{em},ijk}\\
		\frac{1}{\eta_0}\chi_{\text{me},ij} & \chi_{\text{mm},ij} & \frac{1}{2\eta_0k_0}\chi'_{\text{me},ijk} & \frac{1}{2k_0}\chi'_{\text{mm},ijk}\\
		\frac{\epsilon_0}{k_0}Q_{\text{ee},ilj} & \frac{1}{c_0k_0}Q_{\text{em},ilj} & \frac{\epsilon_0}{2k_0^2}Q'_{\text{ee},iljk} & \frac{1}{2c_0k_0^2}Q'_{\text{em},iljk}\\
		\frac{1}{\eta_0k_0}S_{\text{me},ilj} & \frac{1}{k_0}S_{\text{mm},ilj} & \frac{1}{2\eta_0k_0^2 }S'_{\text{me},iljk} & \frac{1}{2k_0^2}S'_{\text{mm},iljk}
	\end{pmatrix}.
\end{equation}
%
Note that we have normalized each tensor in~\eqref{eq:chi} so as to be dimensionless. As can be seen, we retrieve the conventional bianisotropic susceptibility tensors $\te{\chi}_\text{ee}$, $\te{\chi}_\text{mm}$, $\te{\chi}_\text{em}$, and $\te{\chi}_\text{me}$, and a plethora of other terms relating the fields and their gradient to dipolar and quadrupolar responses.

The hypersusceptibility tensor $\te{\chi}$ in~\eqref{eq:PMQS} contains a total number of 576 components. However, several of these components are not independent from each other. Indeed, inspecting the definition of the electric quadrupole moment in~\eqref{eq:Qij} reveals that $Q_{ij} = Q_{ji}$, which directly implies that
\begin{equation}
	Q_{\text{ee},ijk} = Q_{\text{ee},jik}\quad\text{and}\quad Q_{\text{em},ijk} = Q_{\text{em},jik}.
\end{equation}
Additionally, we know that the tensors in~\eqref{eq:chi}, which are related to the derivative of the electric field, are symmetric, as demonstrated in~\eqref{eq:bsplit}. This results in the permutability of their two last indices and thus implies that
\begin{equation}
	\chi'_{\text{ee},ijk} = \chi'_{\text{ee},ikj}\quad\text{and}\quad \chi'_{\text{me},ijk} = \chi'_{\text{me},ikj}.
\end{equation}
The tensor $Q'_{\text{ee},ijkl}$ combines both the symmetry of $Q_{ij}$, which affects its two first indices, and the permutation symmetry of its two last indices due to the fact that it is related to a derivative of the electric field, we thus have that
\begin{equation}
		Q'_{\text{ee},ijkl} = Q'_{\text{ee},jikl} = Q'_{\text{ee},ijlk} = Q'_{\text{ee},jilk}.
\end{equation}
Since there is no symmetry associated with the magnetic quadrupole $\te{S}$ or with the derivative of $\ve{H}$, the tensors $Q'_{\text{em},ijkl}$ and $S'_{\text{me},ijkl}$ only exhibit the partial permutation symmetries
\begin{equation}
	Q'_{\text{em},ijkl} = Q'_{\text{em},jikl}\quad\text{and}\quad S'_{\text{me},ijkl} = S'_{\text{me},ijlk},
\end{equation}	
whereas the tensors $\chi'_{\text{em},ijk}$, $\chi'_{\text{mm},ijk}$, $S_{\text{me},ijk}$, $S_{\text{mm},ijk}$ and $S'_{\text{mm},ijkl}$ exhibit no permutation symmetry at all. Taking into account all of these permutation symmetries, the number of independent components in $\te{\chi}$ is reduced to 420. \B{Note that we are considering that the quadrupolar tensors are expressed in their primitive form implying that the electric quadrupole is symmetric but not traceless and the magnetic quadrupole is traceless but has not been symmetrized. If the irreducible (traceless and symmetrized) moments were considered instead of the primitive ones, then the number of independent elements would be smaller~\cite{raabMultipoleTheoryElectromagnetism2005}. In addition to the symmetries of the multipolar tensors, one should also take into account the symmetries of the material under consideration such as, for instance, the internal atomic symmetries of crystals~\cite{agranovich2013crystal} or, in the case of a metamaterial structure, the symmetries of its constitutive scattering particles~\cite{achouri2021electromagnetic}. These symmetries that are associated with the physical structure itself would further reduce the number of independent hypersusceptibility components in~\eqref{eq:chi}.}

\section{Lorentz Reciprocity Theorem}
\label{sec:lorentz}

\begin{figure}[h]
	\centering
	\includegraphics[width=0.8\columnwidth]{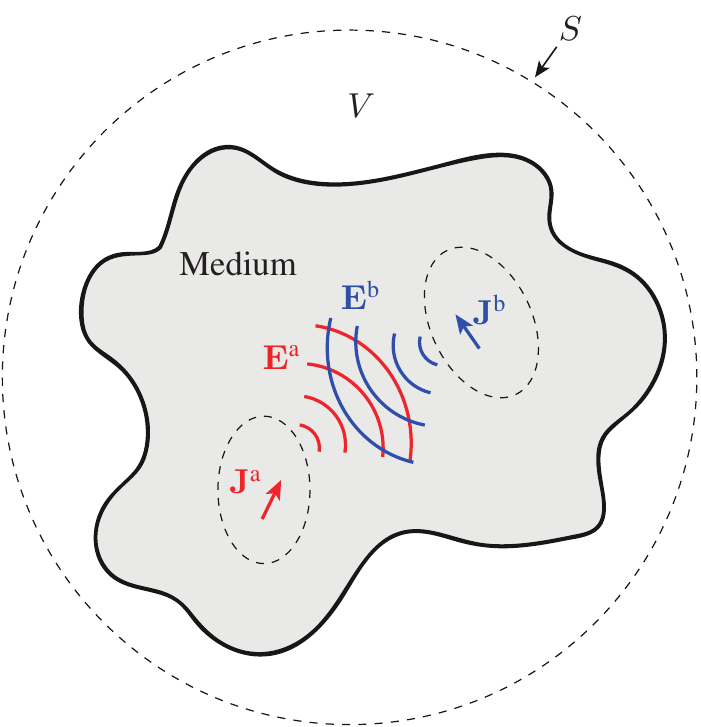}
	\caption{Application of the Lorentz reciprocity theorem showing two regions of a medium, possessing dipolar and quadrupolar responses, interact with each other.}
	\label{fig:lorentz}
\end{figure}

We shall now investigate the reciprocal properties of the tensors $\te{\chi}$ and derive the associated reciprocity conditions that apply to its various subtensors. For this purpose, we start from the definition of electromagnetic reaction given by~\cite{rumseyReactionConceptElectromagnetic1954}
\begin{equation}
	\label{eq:Reaction}
	\langle \text{a}, \text{b}\rangle = \int \ve{E}^\text{a}\cdot\ve{J}^\text{b} dV,
\end{equation}
which results from the interaction of a source \textit{a} with a current distribution $\ve{J}^\text{a}$ producing the field $\ve{E}^\text{a}$ acting, within the volume $V$, on a source \textit{b} with current distribution $\ve{J}^\text{b}$ and electric field $\ve{E}^\text{b}$, as illustrated in Fig.~\ref{fig:lorentz}.

The Lorentz reciprocity theorem then states that the medium contained in $V$ is reciprocal if~\cite{kong1986electromagnetic,rothwell2018electromagnetics,calozElectromagneticNonreciprocity2018}
\begin{equation}
	\label{eq:LorentLemma}
	\langle \text{a}, \text{b}\rangle = \langle \text{b}, \text{a}\rangle.
\end{equation}
Substituting~\eqref{eq:Reaction} into~\eqref{eq:LorentLemma} and using $\ve{J} = \nabla\times\ve{H} - j\omega\ve{D}$ from Maxwell equations along with~\eqref{eq:MEQF1}, yields, after rearranging the terms, the equality~\cite{kong1986electromagnetic,rothwell2018electromagnetics,calozElectromagneticNonreciprocity2018} 
\begin{equation}
	\label{eq:Lorentz}
	\begin{split}
		&\langle \text{a}, \text{b}\rangle - \langle \text{b}, \text{a}\rangle = \int \ve{E}^\text{a}\cdot\ve{J}^\text{b} dV - \int \ve{E}^\text{b}\cdot\ve{J}^\text{a} dV\\	
		&= 	\int\left[j\omega\left(\ve{E}^\text{b}\cdot\ve{D}^\text{a}-\ve{E}^\text{a}\cdot\ve{D}^\text{b}+\ve{H}^\text{a}\cdot\ve{B}^\text{b}-\ve{H}^\text{b}\cdot\ve{B}^\text{a}\right)+\right.\\ &\qquad\qquad\left.	\nabla\cdot\left(\ve{H}^\text{b}\times\ve{E}^\text{a}-\ve{H}^\text{a}\times\ve{E}^\text{b}\right)\right] dV=0
	\end{split}
\end{equation}
This expression, combined with the constitutive relations~\eqref{eq:DBPMQS0}, may now be used to derive the sought after reciprocity conditions. To do so, we next substitute~\eqref{eq:DBPMQS0} into~\eqref{eq:Lorentz} and purposefully ignore the trivial responses related to $\ve{D}=\epsilon_0\ve{E}$ and $\ve{B}=\mu_0\ve{H}$ and only concentrate on the dipolar and quadrupolar terms for convenience. We also use the following identities to simplify the quadrupolar expressions~\cite{silveirinhaBoundaryConditionsQuadrupolar2014}

\B{
\begin{subequations}
	\label{eq:Id}
	\begin{align}
	\ve{E}^\text{u}\cdot\left(\nabla\cdot\te{Q}^\text{v}\right) &= \nabla\cdot\left(\te{Q}^\text{v}\cdot\ve{E}^\text{u}\right) - \sum_i \ve{\hat{u}}_i\cdot\te{Q}^\text{v}\cdot\nabla_i\ve{E}^\text{u},\\
	\ve{H}^\text{u}\cdot\left(\nabla\cdot\te{S}^\text{v}\right) &= \nabla\cdot\left(\te{S}^\text{v}\cdot\ve{H}^\text{u}\right) - \sum_i \ve{\hat{u}}_i\cdot\te{S}^\text{v}\cdot\nabla_i\ve{H}^\text{u},
	\end{align}
\end{subequations}
where $\text{u},\text{v}=\{\text{a},\text{b}\}$ and $\ve{\hat{u}}_1 = \ve{\hat{x}}$, $\ve{\hat{u}}_2 =\ve{\hat{y}}$ and $\ve{\hat{u}}_3 =\ve{\hat{z}}$. Note that the relations in~\eqref{eq:Id} correspond to the application of the Leibniz rule defined as $\nabla_j\left(E_iQ_{ij}\right)=E_i\left(\nabla_jQ_{ij}\right)+\left(\nabla_jE_i\right)Q_{ij}$. }

Next, we apply the divergence theorem to the resulting expression and, after rearranging the terms, obtain
%
%

\B{
\begin{equation}
	\label{eq:LorentzQ2}
	\begin{split}
		&\frac{1}{2}\oint \bigg(2\ve{H}^\text{b}\times\ve{E}^\text{a}-2\ve{H}^\text{a}\times\ve{E}^\text{b}+\\
		&\quad\te{Q}^\text{b}\cdot\ve{E}^\text{a} - \te{Q}^\text{a}\cdot\ve{E}^\text{b}-\mu_0\te{S}^\text{b}\cdot\ve{H}^\text{a} + \mu_0\te{S}^\text{a}\cdot\ve{H}^\text{b}\bigg)\cdot d\ve{S} +\\
		& j\omega\int\left(\ve{E}^\text{b}\cdot\ve{P}^\text{a} -  \frac{1}{2}\sum_i \ve{\hat{u}}_i\cdot\te{Q}^\text{b}\cdot\nabla_i\ve{E}^\text{a} -\right.\\
		&\quad\qquad\left.\ve{E}^\text{a}\cdot\ve{P}^\text{b} +   \frac{1}{2}\sum_i \ve{\hat{u}}_i\cdot\te{Q}^\text{a}\cdot\nabla_i\ve{E}^\text{b} +\right.\\ 
		&\qquad\left.\mu_0\ve{H}^\text{a}\cdot\ve{M}^\text{b}-\frac{\mu_0}{2}\sum_i \ve{\hat{u}}_i\cdot\te{S}^\text{a}\cdot\nabla_i\ve{H}^\text{b} -\right.\\
		&\qquad\left.\mu_0\ve{H}^\text{b}\cdot\ve{M}^\text{a}+\frac{\mu_0}{2}\sum_i \ve{\hat{u}}_i\cdot\te{S}^\text{b}\cdot\nabla_i\ve{H}^\text{a}  \right) dV=0.
	\end{split}
\end{equation}}
Since the surface integral in~\eqref{eq:LorentzQ2} vanishes when taking the surface $S$ to infinity~\cite{kong1986electromagnetic,rothwell2018electromagnetics,calozElectromagneticNonreciprocity2018,silveirinhaBoundaryConditionsQuadrupolar2014}, we next concentrate our attention only on the volume integral. Substituting~\eqref{eq:PMQS} into the integrand of this volume integral, rearranging the terms and keeping only those that are not redundant for briefness, we obtain
%
%
\begin{equation}
	\label{eq:RecipEq}
	\begin{split}
		\int\Big[&\epsilon_0E_i^\text{b}\left(\chi_{\text{ee},ij} - \chi_{\text{ee},ji} \right)E_j^\text{a} -\\
		&\frac{\mu_0}{4k_0^2}\left(S'_{\text{mm},klij}-S'_{\text{mm},ijkl}\right)\nabla_lH_k^\text{a}\nabla_j H_i^\text{b}+ \\
		&\mu_0H_i^\text{a}\left(\chi_{\text{mm},ij} -\chi_{\text{mm},ji}\right)H_j^\text{b} -\\
		&\frac{1}{4c_0k_0^2}\left(Q'_{\text{em},klij} + S'_{\text{me},ijkl}\right)\nabla_lE_k^\text{a}\nabla_j H_i^\text{b}+ \\
		&\frac{\epsilon_0}{2k_0}E_k^\text{b}\left(\chi'_{\text{ee},kji} - Q_{\text{ee},ijk}\right) \nabla_i E_j^\text{a} -\\
		&\frac{1}{2c_0k_0}H_k^\text{b}\left(\chi'_{\text{me},kji}+Q_{\text{em},ijk}\right)\nabla_i E_j^\text{a}+  \\	
		&\frac{1}{2c_0k_0}E_k^\text{b}\left(\chi'_{\text{em},kij} + S_{\text{me},ijk}\right) \nabla_i H_j^\text{a} -\\
		&\frac{\mu_0}{2k_0}H_k^\text{b}\left(\chi'_{\text{mm},kij}-S_{\text{mm},ijk}\right)\nabla_i H_j^\text{a}-  \\	
		&\frac{\epsilon_0}{4k_0^2}\left(Q'_{\text{ee},klij}-Q'_{\text{ee},ijkl}\right)\nabla_lE_k^\text{a}\nabla_j E_i^\text{b} +\\
		&\frac{1}{c_0}E_i^\text{b}\left(\chi_{\text{em},ij} +\chi_{\text{me},ji}\right)H_j^\text{a}\Big] dV= 0.
	\end{split}
\end{equation}
This equation allows us to straightforwardly deduce the reciprocity conditions that the subtensors in $\te{\chi}$ must satisfy. Indeed, since~\eqref{eq:RecipEq} must be zero far any field value, the brackets in~\eqref{eq:RecipEq} must individually be equal to zero. This directly leads to the conventional reciprocity conditions for bianisotropic media, which are given by~\cite{kong1986electromagnetic,rothwell2018electromagnetics,calozElectromagneticNonreciprocity2018}
\begin{equation}
	\label{eq:bianiL}
	\chi_{\text{ee},ij} = \chi_{\text{ee},ji}, \quad \chi_{\text{mm},ij} = \chi_{\text{mm},ji}, \quad \chi_{\text{em},ij} = -\chi_{\text{me},ji}.
\end{equation}
The reciprocity conditions connecting the dipolar susceptibilities, related to field gradients, to the quadrupolar components, related to the fields, are~\cite{LightPropagationCubic1990,simovski2018composite}
\begin{equation}
	\label{eq:3rdL}
\begin{split}
	&\quad\chi'_{\text{ee},kji} = Q_{\text{ee},ijk},\quad  \chi'_{\text{mm},kij} = S_{\text{mm},ijk},\\
	&\chi'_{\text{em},kij} =- S_{\text{me},ijk},\quad \chi'_{\text{me},kji} = -Q_{\text{em},ijk},
\end{split}
\end{equation}
and those applying to the quadrupole components related to the field derivatives read~\cite{LightPropagationCubic1990,simovski2018composite}
\begin{equation}
\label{eq:4thL}
	\begin{split}
	Q'_{\text{ee},klij} &= Q'_{\text{ee},ijkl},\quad Q'_{\text{em},klij} = - S'_{\text{me},ijkl},\\
	&\qquad S'_{\text{mm},klij}=S'_{\text{mm},ijkl}.	
	\end{split}
\end{equation}
Combining all permutation symmetries obtain in Sec.~\ref{sec:SD} with those from reciprocity reduces the number of independent components in $\te{\chi}$ to 210.

\section{Poynting Theorem}
\label{sec:poyn}

We shall now derive the conditions of gainlessness and losslessness applying to the subtensors of $\te{\chi}$ and that can be obtained from the Poynting theorem~\cite{rothwell2018electromagnetics}. The Poynting theorem may be derived by starting from the time-domain Maxwell equation
\begin{subequations}
	\label{eq:MEQ}
	\begin{align}
		\nabla\times\ve{E} &= -\frac{\partial\ve{B}}{\partial t},\label{eq:MEQ1}\\
		\nabla\times\ve{H} &= +\frac{\partial\ve{D}}{\partial t},\label{eq:MEQ2}
	\end{align}
\end{subequations}
and then subtracting~\eqref{eq:MEQ2}, pre-multiplied by $\ve{E}$, to~\eqref{eq:MEQ1}, pre-multiplied by $\ve{H}$, which yields~\footnote{We also use the identity $\nabla\cdot\left(\ve{E}\times\ve{H}\right) = \ve{H}\cdot\left(\nabla\times\ve{E}\right)-\ve{E}\cdot\left(\nabla\times\ve{H}\right)$.}
\begin{equation}
	\label{eq:Poynt}
	\nabla\cdot\ve{S} = -\ve{E}\cdot\frac{\partial\ve{D}}{\partial t} - \ve{H}\cdot\frac{\partial\ve{B}}{\partial t},
\end{equation}
where $\ve{S} = \ve{E}\times\ve{H}$ is the Poynting vector. 

We next develop the two terms on the right-hand side of~\eqref{eq:Poynt}. For brevity, we concentrate our attention only on the first term, which may be split into two equal parts and transformed, using~\eqref{eq:DBPMQS01}, into
\B{
\begin{equation}
	\label{eq:EDdt}
	\ve{E}\cdot\frac{\partial\ve{D}}{\partial t} = \frac{1}{2}\ve{E}\cdot\frac{\partial\ve{D}}{\partial t} + \frac{\epsilon_0}{2}\ve{E}\cdot\frac{\partial\ve{E}}{\partial t}+ \frac{1}{2}\ve{E}\cdot\frac{\partial\ve{P}}{\partial t}- \frac{1}{4}\ve{E}\cdot\frac{\partial}{\partial t}\left(\nabla\cdot\te{Q}\right).
\end{equation}
Next, we add the self-canceling terms $\left(\ve{P}\cdot\partial\ve{E}/\partial t - \ve{P}\cdot\partial\ve{E}/\partial t\right)/2$ and  $[(\nabla\cdot\te{Q})\cdot\partial\ve{E}/\partial t - (\nabla\cdot\te{Q})\cdot\partial\ve{E}/\partial t]/4$ to~\eqref{eq:EDdt} and combine together terms that are similar to each other, to obtain
\begin{equation}
	\label{eq:EDdt2}
	\begin{split}
		\ve{E}\cdot\frac{\partial\ve{D}}{\partial t} &= \frac{1}{2}\frac{\partial}{\partial t}\left(\ve{E}\cdot\ve{D}\right) + \frac{1}{2}\left(\ve{E}\cdot\frac{\partial\ve{P}}{\partial t} - \ve{P}\cdot\frac{\partial\ve{E}}{\partial t}\right) +\\
		&\quad\qquad\frac{1}{4}\left[\left(\nabla\cdot\te{Q}\right)\cdot\frac{\partial\ve{E}}{\partial t} - \ve{E}\cdot\frac{\partial}{\partial t}\left(\nabla\cdot\te{Q}\right)\right]
	\end{split}
\end{equation}}
By the same token, we transform the term $\ve{H}\cdot\partial\ve{B}/\partial t$ in~\eqref{eq:Poynt}, which we then substitute, along with~\eqref{eq:EDdt2}, into~\eqref{eq:Poynt} yielding
\B{
\begin{equation}
	\label{eq:PT}
	\begin{split}
			\frac{\partial w}{\partial t} + &\nabla\cdot\ve{S} = -\frac{1}{2}\left(\ve{E}\cdot\frac{\partial\ve{P}}{\partial t} - \ve{P}\cdot\frac{\partial\ve{E}}{\partial t}\right)-\\
			&\frac{1}{4}\left[\left(\nabla\cdot\te{Q}\right)\cdot\frac{\partial\ve{E}}{\partial t} - \ve{E}\cdot\frac{\partial}{\partial t}\left(\nabla\cdot\te{Q}\right)\right]-\\
			&\frac{\mu_0}{2}\left(\ve{H}\cdot\frac{\partial\ve{M}}{\partial t} - \ve{M}\cdot\frac{\partial\ve{H}}{\partial t}\right)-\\
			&\frac{\mu_0}{4}\left[\left(\nabla\cdot\te{S}\right)\cdot\frac{\partial\ve{H}}{\partial t} - \ve{H}\cdot\frac{\partial}{\partial t}\left(\nabla\cdot\te{S}\right)\right],
	\end{split}
\end{equation}}
where
\begin{equation}
	\label{eq:w}
	w = \frac{1}{2}\left(\ve{E}\cdot\ve{D}+\ve{H}\cdot\ve{B}\right).
\end{equation}
Relation~\eqref{eq:PT} along with~\eqref{eq:w} constitute the instantaneous Poynting theorem in the presence of electric and magnetic dipoles and quadrupoles.

We next assume time-harmonic fields, which transforms the time derivatives in~\eqref{eq:PT} into $\partial/\partial t \rightarrow j\omega$, and take the time-average counterpart\B{~\footnote{\B{We have used the following procedure to obtain the time-average counterpart of~\eqref{eq:PT}. Consider the time-domain equation $\ve{A}(t)=\ve{B}(t)\cdot\ve{C}(t)$, where $\ve{B}(t)$ and $\ve{C}(t)$ are time-harmonic vectors that can be expressed as $\ve{B}(t)=\operatorname{Re}\left(\ve{B}e^{j\omega t}\right)$ and $\ve{C}(t)=\operatorname{Re}\left(\ve{C}e^{j\omega t}\right)$. The time-average counterpart of $\ve{A}(t)$ is now obtained as $\langle\ve{A}\rangle = \frac{1}{2}\operatorname{Re}\left(\ve{B}^*\cdot\ve{C}\right)$}}} of~\eqref{eq:PT} to obtain
\B{
\begin{equation}
	\label{eq:PTav}
	\begin{split}
		&\nabla\cdot\langle\ve{S}\rangle = \frac{\omega}{8}\operatorname{Im}\bigg[2\ve{E}^*\cdot\ve{P} - 2\ve{P}^*\cdot\ve{E}+\left(\nabla\cdot\te{Q}\right)^*\cdot\ve{E}-\\
		&\qquad \ve{E}^*\cdot\left(\nabla\cdot\te{Q}\right)+2\mu_0\left(\ve{H}^*\cdot\ve{M} - \ve{M}^*\cdot\ve{H}\right)+\\
		&\qquad\qquad\mu_0\left(\nabla\cdot\te{S}\right)^*\cdot\ve{H} - \mu_0\ve{H}^*\cdot\left(\nabla\cdot\te{S}\right)\bigg],
	\end{split}
\end{equation}
where $\langle\partial w/\partial t\rangle = 0$. Finally, we integrate~\eqref{eq:PTav} over a volume $V$ and simplify the resulting expression using the identities~\eqref{eq:Id} with  $\text{u},\text{v}=\{^*,\}$, e.g. $\text{u}$ corresponds to the complex conjugate operation while $\text{v}$ performs no operation,  and vice versa and, noting that $(\nabla\cdot\te{A})^*=\nabla\cdot\te{A}^*$, with $\te{A}=\{\te{S}, \te{Q}\}$, we obtain 
}

\begin{figure}[h!]
	\centering
	\includegraphics[width=0.8\columnwidth]{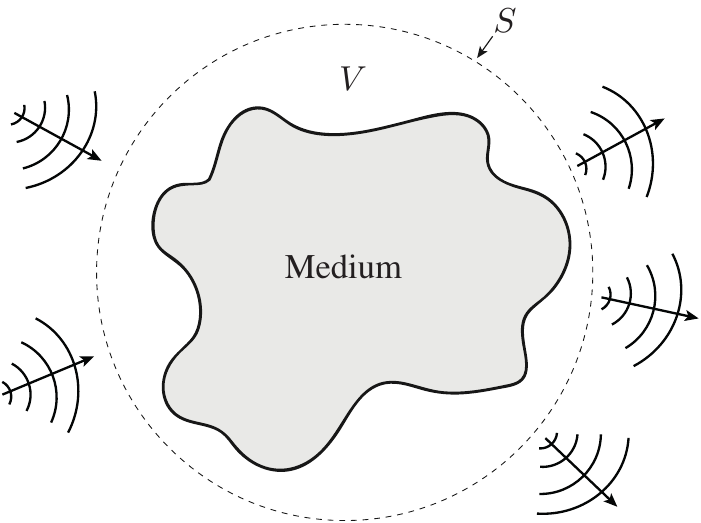}
	\caption{Application of the Poynting theorem showing electromagnetic waves impinging on and being scattered by a lossless and gainless medium possessing dipolar and quadrupolar responses. Energy conservation requires that the energy entering $V$ equals the energy leaving it.}
	\label{fig:Poynt}
\end{figure}

\B{
\begin{equation}
	\label{eq:PTav2}
	\begin{split}
		&\oint\langle\ve{S}\rangle\cdot d\ve{S} = \frac{\omega}{8}\operatorname{Im}\bigg[\oint \Big(\te{Q}^*\cdot\ve{E} - \te{Q}\cdot\ve{E}^* + \\
		&\qquad\mu_0\te{S}^*\cdot\ve{H} - \mu_0\te{S}\cdot\ve{H}^*\Big)\cdot d\ve{S}+\\		
		&\int\bigg(~ 2\ve{E}^*\cdot\ve{P}   - \sum_i \ve{\hat{u}}_i\cdot\te{Q}^*\cdot\nabla_i\ve{E}-\\
		&\qquad~2\ve{E}\cdot\ve{P}^*         +\sum_i \ve{\hat{u}}_i\cdot\te{Q}  \cdot\nabla_i\ve{E}^*+\\
		&\quad2\mu_0\ve{H}^*\cdot\ve{M} -\mu_0\sum_i \ve{\hat{u}}_i\cdot\te{S}^*\cdot\nabla_i\ve{H}-\\
		&\quad2\mu_0\ve{H}\cdot\ve{M}^* +\mu_0\sum_i \ve{\hat{u}}_i\cdot\te{S}  \cdot\nabla_i\ve{H}^*\bigg)dV\bigg].
	\end{split}
\end{equation}}
Now, for the same reason mentioned in Sec.~\ref{sec:lorentz}, the surface integral on the right-hand side of~\eqref{eq:PTav2} vanishes when $S$ is taken to infinity. The conditions of gainlessness and losslessness are now obtained from the integrand of the volume integral in~\eqref{eq:PTav2} knowing that it must be zero since, to satisfy energy conservation, all the energy entering the volume $V$ must leave it, i.e., $\oint\langle\ve{S}\rangle\cdot d\ve{S}=0$, as illustrated in Fig.~\ref{fig:Poynt}. 

After substituting~\eqref{eq:PMQS} into~\eqref{eq:PTav2} and selecting the terms that must cancel each other as in~\eqref{eq:RecipEq}, we obtain the desired conditions for bianisotropic media as~\cite{serdkov2001electromagnetics,rothwell2018electromagnetics}
\begin{equation}
	\label{eq:bianiP}
	\chi_{\text{ee},ij} = \chi^*_{\text{ee},ji}, \quad \chi_{\text{mm},ij} = \chi^*_{\text{mm},ji}, \quad \chi_{\text{em},ij} = \chi^*_{\text{me},ji}.
\end{equation}
The conditions applying to the third-rank tensors are then given by
\begin{subequations}
	\label{eq:3rdP}
	\begin{align}
		&\quad\chi'_{\text{ee},kji} = Q^*_{\text{ee},ijk},\quad  \chi'_{\text{mm},kij} = S^*_{\text{mm},ijk}\\
		&\chi'_{\text{em},kij} = S^*_{\text{me},ijk},\quad \chi'_{\text{me},kji} = Q^*_{\text{em},ijk},
	\end{align}
\end{subequations}
whereas those for the fourth-rank tensors read
\begin{equation}
	\label{eq:4thP}
	\begin{split}
	&Q'_{\text{ee},klij} = Q'^*_{\text{ee},ijkl},\quad S'_{\text{mm},klij}=S'^*_{\text{mm},ijkl},\\
	&\qquad\qquad Q'_{\text{em},klij} =  S'^*_{\text{me},ijkl}.
	\end{split}
\end{equation}
Comparing relations~\eqref{eq:bianiP},~\eqref{eq:3rdP} and~\eqref{eq:4thP} to the reciprocity conditions~\eqref{eq:bianiL},~\eqref{eq:3rdL} and~\eqref{eq:4thL} reveals that these two sets of conditions are almost identical to each other. Indeed, the relations deduced from the Poynting theorem do not imply additional permutation symmetries compared to those already provided by the reciprocity conditions. Instead, they require some tensors to be purely real and some others to be purely imaginary for the medium to satisfy energy conservation.

\section{Conclusions}
\label{sec:concl}

We have provided a self-consistent and purely electromagnetic derivation of the Lorentz and Poynting theorems, and have deduced the associated conditions of reciprocity and gainlessness and losslessness, in the presence of electric and magnetic quadrupolar responses expressed in terms of fields and field gradients. We expect that these conditions will be especially useful for the developments of advanced metamaterial and metasurface modeling techniques requiring the presence of quadrupolar responses.

\section*{Acknowledgements}

We gratefully acknowledge funding from the European Research Council (ERC-2015-AdG-695206 Nanofactory).

\bibliography{NewLib}

\end{document}